\documentclass[
    aps, prd, nofootinbib, twocolumn, floatfix, superscriptaddress, secnumarabic,
    USenglish 
]{revtex4-2}

\usepackage{
    babel,
    amsfonts, amssymb, amsmath, amstext, dsfont,
    graphicx, subfigure,
    placeins, float,
    physics, slashed,
    booktabs,
}
\usepackage[svgnames]{xcolor}

\newcommand{\Z}{\mathcal{Z}}
\newcommand{\D}{\mathcal{D}}
\newcommand{\ZT}{\mathrm{Z}(3)}
\newcommand{\SUT}{\mathrm{SU}(3)}

\usepackage[
    bookmarks,
    bookmarksopen = true,
    bookmarksnumbered = true,
    breaklinks = true,
    linktocpage,
    colorlinks = true,
    linkcolor = DarkRed,
    urlcolor  = blue,
    citecolor = Navy,
    anchorcolor = green,
    hyperindex = true,
    hyperfigures
]{hyperref}

\usepackage[nameinlink]{cleveref} 

\crefname{figure}{fig.}{figs.}%

\makeatletter
    \crefformat{equation}{#2\cref@equation@name\,(#1)#3}
    \labelcrefformat{equation}{#2(#1)#3}
\makeatother

\let\oldcite\cite
\renewcommand{\cite}[1]{\mbox{\oldcite{#1}}}

\usepackage{xstring}
\newcommand{\referencename}{Ref.}
\newcommand{\referencenames}{Refs.}
\newcommand{\refcite}[1]{%
  \IfSubStr{#1}{,}
    {\referencenames\,\cite{#1}}%
    {\referencename\,\cite{#1}}%
}

\DeclareFontFamily{U}{matha}{}
\DeclareFontShape{U}{matha}{m}{n}{
  <-5.5>    matha5
  <5.5-6.5> matha6 
  <6.5-7.5> matha7
  <7.5-8.5> matha8
  <8.5-9.5> matha9
  <9.5-11>  matha10
  <11->     matha12
}{}
\DeclareSymbolFont{matha}{U}{matha}{m}{n}
\DeclareFontSubstitution{U}{matha}{m}{n}
\DeclareFontFamily{U}{mathx}{\hyphenchar\font45}
\DeclareFontShape{U}{mathx}{m}{n}{<-> mathx10}{}
\DeclareSymbolFont{mathx}{U}{mathx}{m}{n}
\DeclareFontSubstitution{U}{mathx}{m}{n}
\makeatletter
\DeclareMathSymbol{\rightarrow}{3}{matha}{"D1}
\DeclareMathSymbol{\to}{3}{matha}{"D1}
\makeatother

\graphicspath{{anc/}}

\begin{document}

\title{Exact center symmetry and first-order phase transition in QCD with three degenerate dynamical quarks}

\author{G.\ Endr\H{o}di}

\affiliation{Institute of Physics and Astronomy,
ELTE E\"otv\"os Lor\'and University,\\
P\'azm\'any P.\ s\'et\'any 1/A, H-1117 Budapest, Hungary}

\author{G.\ D.\ Moore}

\affiliation{Institut f\"ur Kernphysik, Technische Universit\"at Darmstadt,\\ Schlossgartenstra{\ss}e 2,
D-64289 Darmstadt, Germany}

\author{A.\ Pieczynski}

\affiliation{Fakult\"at f\"ur Physik, Universit\"at Bielefeld,\\ Universit\"{a}tsstra{\ss}e 25, D-33615 Bielefeld, Germany}

\affiliation{Fakult\"at f\"ur Mathematik und Naturwissenschaften,Universität Wuppertal, Gau\ss str.\ 20, 42119 Wuppertal, Germany}

\author{A.\ Sciarra}

\affiliation{Institut für Theoretische Physik, Goethe Universität Frankfurt, Max-von-Laue-Str.\ 1, 60438 Frankfurt, Germany}


\begin{abstract}
We study QCD with three degenerate flavors of dynamical quarks using first-principles lattice simulations. 
For a specific choice of imaginary isospin chemical potential, this theory possesses an exact center symmetry, just like pure gauge theory.
This exact symmetry is expected to be intact at low temperatures and spontaneously broken in the high-temperature regime.
By analyzing the finite-size scaling of the Polyakov loop distribution, obtained with a dedicated multi-histogram approach,
we demonstrate that there is a first-order deconfinement phase transition in between. Our results are obtained 
employing stout-smeared rooted staggered quarks at one lattice spacing. Using simulations at different quark masses we sketch the behavior of QCD in the mass-isospin chemical potential plane, shedding new light on this corner of the fundamental phase diagram of the strong interactions and the relationship between chiral symmetry breaking and deconfinement.
\end{abstract}

\maketitle

\section{Introduction}
\label{Introduction}

Quantum Chromodynamics (QCD) is the theory of the strong interactions, exhibiting at least two different phases -- a deconfined, chirally symmetric phase at high energies and a confined phase with spontaneously broken chiral symmetry in the low-energy regime. The nature of the transition between these phases is relevant for the cosmological evolution of the early Universe as well as for high-energy heavy-ion collision phenomenology~\cite{Aarts:2023vsf}.

At high temperature but zero density, the QCD transition at the physical point is an analytical crossover~\cite{Aoki:2006we,Bhattacharya:2014ara}.
It is known that the nature of the transition is sensitive to the masses of the light and strange quarks,
which is described by the so-called Columbia plot~\cite{Brown:1990ev}.
Towards large quark masses, the first-order phase transition of pure gauge theory~\cite{Yaffe:1982qf,Svetitsky:1982gs}
is approached, where the $\ZT$ center symmetry associated with (de)confinement becomes exact, see \refcite{Borsanyi:2022xml} for the most recent result. This behavior persists for sufficiently heavy quarks, see e.g.\ \refcite{Ejiri:2019csa,Cuteri:2020yke,Kiyohara:2021smr}.

In the opposite limit of massless quarks, chiral symmetry is the relevant issue. The chiral limit is significantly more intricate,
and there are several possible scenarios for the order of the transition there, although recent studies find indications for a second-order phase transition for two and three massless quark flavors~\cite{HotQCD:2019xnw,Cuteri:2021ikv,Dini:2021hug}.
The structure of the Columbia plot is in general expected
to be relevant for constraining the behavior of QCD at nonzero quark density and the associated phase diagram
in the plane spanned by the temperature $T$ and the 
baryon chemical potential $\mu_B$.

The QCD action at $\mu_B\neq0$ is complex, hindering standard,
importance-sampling-based simulation algorithms. 
Currently, one of the most popular approaches to overcome the complex action problem and to study 
dense QCD is by means of an analytical continuation of results
obtained for imaginary chemical potentials, where the 
action is real and direct simulations are amenable~\cite{DElia:2002tig,deForcrand:2002hgr}.
In addition, QCD at imaginary chemical potentials exhibits 
a rich phase structure at high temperatures that can be studied already via leading-order perturbation theory~\cite{PhysRevD.24.475,PhysRevD.25.2667,Roberge:1986mm}.
In particular, the $T-i\mu_B$ plane features first-order Roberge-Weiss phase transitions at specific values of the imaginary baryon chemical potential, where a $\mathrm{Z}(2)$ subgroup of the center symmetry of pure gauge theory is recovered.

The complete description of the dense three-flavor theory involves three chemical potentials and besides the baryon chemical potential it is customary to 
use the strangeness and isospin quantum numbers, conjugate to the strangeness chemical potential $\mu_S$ and the isospin chemical potential $\mu_I$. The latter is special in the sense
that QCD at real values of $\mu_I$ is free of the complex action problem. The associated phase diagram~\cite{Brandt:2017oyy,Brandt:2018omg} and the equation of state~\cite{Brandt:2022hwy,Abbott:2023coj} have been determined by first-principles lattice simulations.
The equation of state of the theory at nonzero isospin is also relevant because it may provide non-trivial constraints for 
the baryonic equation of state, necessary for neutron star physics~\cite{Moore:2023glb,Fujimoto:2023unl}.

Recently, the Roberge-Weiss transitions present at $i\mu_B\neq0$ have been generalized to the
two-flavor theory and the leading perturbative behavior was explored in the $i\mu_B-i\mu_I$ plane~\cite{Brandt:2022iwk,Brandt:2022jwo}. Interestingly, at specific points on this plane, the complete $\ZT$ center symmetry is
recovered, although the location of the associated phase transitions was found to depend on the temperature.
In contrast, the three-flavor theory (with degenerate quark masses) is known to be $\ZT$-symmetric at the
special value of $i\mu_I/T=4\pi/3$ and $\mu_B=\mu_S=0$ for any temperature~\cite{Kouno:2012zz,Cherman:2017tey}.
This theory has been studied extensively in terms of model calculations~\cite{Kouno:2012zz,Kouno:2013zr,Kouno:2015sja,Hirakida:2017bye,Li:2018xgl,Kouno:2021eou}, perturbation theory~\cite{HIRAKIDA:2020ljz,Poppitz:2021cxe} as well as lattice simulations~\cite{Iritani:2015ara,Misumi:2015hfa}.
The exact center symmetry naturally suggests that -- just like in pure gauge theory -- the high-temperature QCD transition may be first order in this setup -- an expectation backed up by the above approaches. However, a fully non-perturbative study that explores the finite-size scaling behavior near the transition is currently missing.

In this paper, we aim to fill this gap.
First, using the one-loop perturbative effective action of the Polyakov loop, we determine the phase diagram of the three-flavor theory in the $i\mu_B-i\mu_I-i\mu_S$ space at very high temperatures and identify the center-symmetric points.
Focusing on the special choice of $i\mu_I/T=4\pi/3$ and $\mu_B=\mu_S=0$ mentioned above, we perform full lattice simulations using three mass-degenerate stout-smeared rooted staggered quarks on $N_t=6$ lattices. Through a finite-size scaling analysis, facilitated by an optimized multi-histogram reweighting approach, we demonstrate that a first-order phase transition indeed occurs as the temperature is reduced. Our main results are obtained at the physical strange quark mass but we also explore different quark mass values and argue that the main conclusion persists for any massive theory. Finally, we sketch the generalized phase diagram of QCD in the imaginary isospin chemical potential -- mass -- temperature space.
Our preliminary results have been presented in \refcite{Endrodi:2025hlb}.

\section{Center symmetry and perturbative effective potential}

We consider QCD at nonzero temperature $T$ and finite volume $V$ with three dynamical quark flavors sharing the same mass $m$. To control the density of each quark species, the chemical potentials $\mu_f$ are introduced, where $f=u,d,s$ labels the flavors. The basis spanned by the baryon, isospin and strangeness quantum numbers is given by
\begin{equation}
\mu_u=\frac{\mu_B}{3}+\frac{\mu_I}{2}, \quad
\mu_d=\frac{\mu_B}{3}-\frac{\mu_I}{2}, \quad
\mu_s=\frac{\mu_B}{3}-\mu_S\,.
\label{eq:chempotbasis}
\end{equation}
The partition function of the system is given by the Euclidean path integral over the gluon field $A_\nu$,
\begin{equation}
\Z = \int \D A_\nu\, e^{-S_g}\! \prod_{f=u,d,s} \det [\slashed{D}(A_\nu + i\mu_f\delta_{\nu4})+m]\,,
 \label{eq:Z1}
\end{equation}
where $S_g=S_g[A_\nu]$ is the gauge action. The theory is periodic in the imaginary quark chemical potentials $i\mu_f$ with period $2\pi T$%
\footnote{As discussed in the \cref{app:pert}.\label{fn:1}, the coordinates \mbox{$0\le i\mu_B< 6\pi T$}, \mbox{$0\le i\mu_I<4\pi T$}, \mbox{$0\le i\mu_S<2\pi T$} cover this region twice.}.

In the $m\to\infty$ limit, quarks decouple and pure gauge theory is recovered, which is invariant under center transformations. The latter are 
large gauge transformations, i.e.\ gauge transformations
\begin{equation}
 A_\nu(x)\to \Omega(x)  ( A_\nu(x)+i\partial_\nu) \Omega^\dagger(x)\,,
 \label{eq:center1}
\end{equation}
satisfying the twisted boundary condition
\begin{equation}
 \Omega(x_4=1/T) = z\, \Omega(x_4=0)\,,
 \label{eq:center2}
\end{equation}
where $z$ is a center element
\begin{equation}
 z\in\ZT=\{\mathds{1},e^{i2\pi/3}\mathds{1},e^{-i2\pi/3}\mathds{1}\}\subset\SUT\,.
 \label{eq:center3}
\end{equation}
Center symmetry is intact at low temperature and spontaneously broken at high $T$.  The associated order parameter is the Polyakov loop,
\begin{equation}
 P = \frac{1}{3V}\int\dd^3 \bm x\, \Tr \mathcal{P} \exp\left[ i \int^{1/T}_0 \mathrm{d}x_4 A_4 (\bm x, x_4)\right]\,,
 \label{eq:ploopdef1}
\end{equation}
which transforms as $P\to z P$ under a center transformation and assumes nonzero values in the deconfined phase. In \cref{eq:ploopdef1}, $\mathcal{P}$ denotes the path-ordering operation.
In the presence of dynamical quarks, center symmetry is explicitly broken, since center transformations affect the time-like boundary conditions of fermions, modifying the quark determinants in \cref{eq:Z1}. The Polyakov loop still acts as an approximate order parameter, assuming small values at low and large values at high $T$.
For a pedagogical introduction on center symmetry and the Polyakov loop, see e.g.\ \refcite{Fukushima:2017csk}.

At very high temperatures, asymptotic freedom ensures that QCD is weakly coupled, enabling a perturbative treatment of this theory. To determine the effective potential $V_{\rm eff}(P)$ for the Polyakov loop~\labelcref{eq:ploopdef1}, one needs to consider homogeneous time-like color background fields $A_4$. To one-loop order, this can be carried out in a closed (semi-)analytic form. As visible from \cref{eq:Z1}, the imaginary chemical potential merely amounts to a shift of the background color field. This observation allows one to determine the phase diagram of the theory as a function of the chemical potentials, as was done first in the seminal papers by Roberge and Weiss~\cite{PhysRevD.24.475,PhysRevD.25.2667,Roberge:1986mm} for a single quark flavor. The minima of the effective potential correspond to the so-called center sectors with $\arg P= 2\pi k/3$ with $k\in\{0,1,2\}$ and are separated by first-order phase transitions as $i\mu_f$ are varied.

\begin{figure}[tb]
  \includegraphics[width=.9\columnwidth]{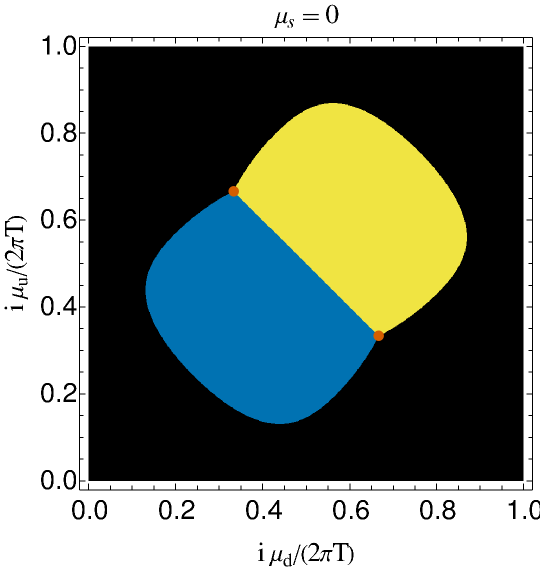}
 \caption{\label{fig:pertpd}
Phase diagram of QCD with three massless quark flavors in the $i\mu_u-i\mu_d$ plane at $\mu_s=0$.
The colors indicate the three center sectors, where the Polyakov loop phase is real (black) positive imaginary (yellow/light) or negative imaginary (blue/medium).
The red dots indicate $\ZT$-symmetric points.
Opposite edges of the diagram are identified due to the periodicity in the imaginary chemical potentials.
The case of a pure isospin chemical potential is described by points along the diagonal from the top-left to the bottom-right corner.
One of the two triple points matches the triple point identified in \cref{eq:imagiso_choice} after shifting the negative-valued chemical potential up by $2 \pi T$ and the second one is a result of exchanging the up- and down-values.} 
\end{figure}

In the \cref{app:pert}, we generalize the perturbative analysis to three quark flavors, following the two-flavor study of \refcite{Brandt:2022jwo}, and determine the effective potential as a function of the imaginary chemical potentials, $V_{\rm eff}(P;i\mu_u,i\mu_d,i\mu_s)$. The resulting phase diagram for $\mu_s=0$ is shown in \cref{fig:pertpd} for massless quarks. Here one can observe the non-trivial shape of the three center sectors and the first-order phase transitions between them. Along the lines separating two different sectors, the theory is $\mathrm{Z}(2)$-symmetric, while at the meeting point of three sectors, $\ZT$-symmetry is realized. 

One of the $\ZT$-symmetric points is at 
$i\mu_I=4\pi T/3$ and $\mu_B=\mu_S=0$, corresponding to the 
the quark chemical potentials
\begin{equation}
i\mu_u=-i\mu_d=2\pi T/3, \qquad i\mu_s=0\,.
\label{eq:imagiso_choice}
\end{equation}
Even if each fermion determinant in \cref{eq:Z1} is modified by the center transformation~\labelcref{eq:center1,eq:center2,eq:center3}, in this case
the product of fermion determinants remains invariant, as the transformation merely corresponds to a permutation of the flavors. This invariance is
therefore sometimes called color-flavor-center symmetry~\cite{Kouno:2012zz}.
The other center-symmetric points can be found from this one via charge conjugation symmetry, see also \refcite{Brandt:2022jwo}. Further details of the effective potential and the three-dimensional phase diagram are discussed in the \cref{app:pert}.
We note that the general conditions for exact center symmetry for $N_c$ colors and $N_f$ flavors were discussed in \refcite{Cherman:2017tey}.

\section{Lattice setup and methods}

Next, we specialize to the choice~\labelcref{eq:imagiso_choice} of the imaginary chemical potentials. 
To simulate the path integral~\labelcref{eq:Z1}, we consider the link variables $U_\nu=e^{iA_\nu}\in\SUT$ and use the rooted staggered formalism.
Specifically, $S_g$ is discretized using the tree-level improved Symanzik action and $\slashed{D}$ using the staggered Dirac operator with twofold stout-smeared~\cite{Morningstar:2003gk} links $\widetilde U_\nu(n)$ obtained from the unsmeared links $U_\nu(n)$.

We employ lattices with geometry $N_s^3\times N_t$ and the sites will be labeled below by the four integers $\{n_1,n_2,n_3,n_4\}\equiv\{\mathbf{n},n_4\}$. The temperature and the spatial volume are given by $T=1/(N_t\,a)$ and $V=(N_s\,a)^3$, where $a$ is the lattice spacing. 
Accordingly, the temperature is changed by tuning the inverse bare coupling $\beta$ and the lattice scale $a(\beta)$ is set at the physical point, corresponding to a mass-independent renormalization scheme.
The masses of all three quarks are set to the physical strange quark mass and run as a function of $\beta$ along the line of constant physics, which is again determined at the physical point~\cite{Borsanyi:2010cj}.
The three Dirac operators only differ in the imaginary chemical potentials $i\mu_f$. These enter the temporal hoppings in an exponential form, $\widetilde U_4(n)\cdot e^{ia\mu_f}$.

On the level of the gluon links, the center transformation~\labelcref{eq:center1,eq:center2,eq:center3} is realized by
\begin{equation}
\forall\, \mathbf{n}:\qquad U_4(\mathbf{n},n_4)\to z\,U_4(\mathbf{n},n_4),
 \label{eq:Z3}
\end{equation}
at a given time-slice $n_4$.
The above transformation implies the same replacement for the smeared links $\widetilde U_4$, too. Using \cref{eq:Z3} and the fact that the links multiply $e^{ia\mu_f}$ in the Dirac operators, the center symmetry of the theory is transparent: a transformation with for example $z=e^{i2\pi/3}\mathds{1}$ merely 
interchanges the quarks as $u\to d$, $d\to s$ and $s\to u$. Since all three flavors share the same mass, this is indeed an invariance of the theory.

Using the gluon links, the Polyakov loop~\labelcref{eq:ploopdef1} takes the form,
\begin{equation}
 P=\frac{1}{3V}\sum_\mathbf{n}\Tr \prod_{n_4=0}^{N_t-1} \widetilde U_4(\mathbf{n},n_4)\,,
 \label{eq:Pdef}
\end{equation}
transforming as $P\to z P$ under~\labelcref{eq:Z3}.
Note that we use the stout-smeared links $\widetilde U_4$ to construct this operator in order to suppress noise. The symmetry, being exact, forces the expectation value of the Polyakov loop to vanish at any temperature -- just like in pure gauge theory. To characterize the transition via expectation values involving the Polyakov loop operator, one therefore needs an alternative observable. 

One possibility is to consider the `rotated' Polyakov loop $\bar P$~\cite{Iwasaki:1992ik}, constructed as follows. For any gauge field configuration $U$, the phase $\arg P$ of the Polyakov loop is calculated. Next, the Polyakov loop is rotated to the real sector via the center transformation $z_U\in\mathrm{Z}(3)$ chosen such that $z_U P$ is in the real sector. The rotated Polyakov loop is the real part of this complex number,
\begin{equation}
 \bar P = \Re \left(z_U P\right), \qquad
 -\pi/3 < \arg \left(z_U P\right) \le \pi/3\,.
 \label{eq:Pbardef}
\end{equation}
For the so defined observable, $\langle \bar P \rangle \neq0$. It is expected to give a clear signal of the transition~\cite{Iwasaki:1992ik}.

In addition, we also discuss higher central moments $\kappa_n \equiv \langle (\bar P - \langle \bar P \rangle)^n\rangle$ of the distribution of the rotated Polyakov loop. The second, third and fourth moments define the susceptibility, the skewness and the kurtosis of $\bar P$,
\begin{equation}
 \chi_{\bar P} = V \kappa_2, \qquad
 s_{\bar P} = \frac{\kappa_3}{\kappa_2^{3/2}}, \qquad
 B_{\bar P} = \frac{\kappa_4}{\kappa_2^2}\,,
 \label{eq:Pbarmoments}
\end{equation}
which are constructed to be intensive observables.

Finally, to assess the breaking and restoration of chiral symmetry, we consider the sum of quark condensates,
\begin{equation}
 \bar\psi\psi = \frac{1}{4} \frac{T}{V}\sum_{f=u,d,s} \Tr \frac{1}{\slashed{D}(\mu_f)+m}\,,
 \label{eq:condensate}
\end{equation}
which is also invariant under center transformations~\cite{Cherman:2017tey}.

On our $N_t=6$ ensembles, we consider inverse gauge couplings \mbox{$3.79\le \beta\le 3.85$}, corresponding to temperatures \mbox{$232\textmd{ MeV}\le T\le 333\textmd{ MeV}$}. To investigate the finite volume scaling of our observables, we simulate lattices with \mbox{$N_s\in\{16, 20, 24, 28, 32\}$}. 
To explore the impact of changing the quark masses, we perform further runs on $16^3\times6$ lattices in a broader range of $\beta$ values.
Moreover, for scale setting purposes we also simulated $24^4$ ensembles, which are at approximately zero temperature.

\subsection{Multi-histogram method}
\label{sec:mh}

In order to locate the critical temperature and find the scaling of the above observables with the volume, an efficient interpolation method in $\beta$ is necessary. To this end we employ the multi-histogram method~\cite{Ferrenberg:1989ui,newmanb99}, which allows for an effective weighted averaging of reweightings from each of our simulation points. For this work, \texttt{v0.3} of our publicly available ``Monte Carlo {\texttt{C++}} analysis tools''~\cite{MCC++} has been used.

Furthermore, the different simulation points -- to be combined using the multi-histogram method -- differ not only in $\beta$ but also in $m$ due to the tuning along our line of constant physics $m(\beta)$. This necessitates the reweighting in $\beta$ to be accompanied with a reweighting in $m$. A complete reweighting in the quark mass would require the calculation of the fermion determinant and is therefore no option. Instead, we expand the corresponding reweighting factor in the quark mass. Specifically, a leading-order expansion of the logarithm of the reweighting factor is done. Therefore, the complete reweighting factor involved in a reweighting from $\beta$ to $\beta'$ takes the form,
\begin{equation}
 W(\beta\to\beta') \approx \exp\!\left[ - (\beta'-\beta) \!\left( S_{g} - \frac{\partial m}{\partial \beta}\frac{V}{T}\, \bar\psi\psi\right) \right],
 \label{eq:rewfac}
\end{equation}
where $S_{g}$ is the gauge action and  $\bar\psi\psi$ the quark condensate~\labelcref{eq:condensate}. The scale function $\partial m/\partial \beta$ is known from the line of constant physics determination~\cite{Borsanyi:2010cj}.
The leading-order expansion is expected to be a good approximation to the reweighting factor as long as $|\beta'-\beta|$ is small. Since our primary objective is the investigation of observables in a narrow interval around the critical point, this approach is justified. 

The reweighting factor~\labelcref{eq:rewfac} suffices to determine the expectation values of operators that do not depend explicitly on $\beta$ or $m$, like the (rotated) Polyakov loop,
\begin{equation}
\langle \bar{P} \rangle_{\beta'} = \frac{\langle \bar P \,W(\beta\to\beta') \rangle_{\beta} }{ \langle W(\beta\to\beta')\rangle_{\beta}    }\,.
\end{equation}
For the quark condensate~\labelcref{eq:condensate}, an explicit dependence on $m$ is present, requiring the evaluation of the operator at the shifted quark mass $m(\beta')$. In particular, the leading-order reweighting involves the leading-order expansion of this dependence,
\begin{equation}
 \bar\psi\psi(\beta') \approx \bar\psi\psi(\beta) + (\beta'-\beta)\cdot\frac{\partial m}{\partial \beta}\,\chi_{\bar\psi\psi}(\beta)\,,
\end{equation}
involving the connected susceptibility,
\begin{equation}
 \chi_{\bar\psi\psi} = \frac{\partial \bar\psi\psi}{\partial m} = -\frac{1}{4}\frac{T}{V} \sum_{f=u,d,s} \Tr\frac{1}{ \left[\slashed{D}(\mu_f)+m\right]^{2}}\,.
\end{equation}
Thus, the reweighting of the condensate altogether takes the form
\begin{align}
\langle \bar\psi\psi(\beta') \rangle_{\beta'} &= \frac{\langle W(\beta\to\beta') \, \bar\psi\psi(\beta)  \rangle_{\beta} }{ \langle W(\beta\to\beta')\rangle_{\beta}} \\
&+ (\beta'-\beta)\cdot\frac{\partial m}{\partial \beta}
\frac{\langle W(\beta\to\beta') \, \chi_{\bar\psi\psi}(\beta) \rangle_{\beta} }{ \langle W(\beta\to\beta')\rangle_{\beta}    }\,.\notag
\end{align}

\section{Results}

\subsection{Polyakov loop}

\begin{figure}[t]
 \centering
 \includegraphics[height=0.275\textheight]{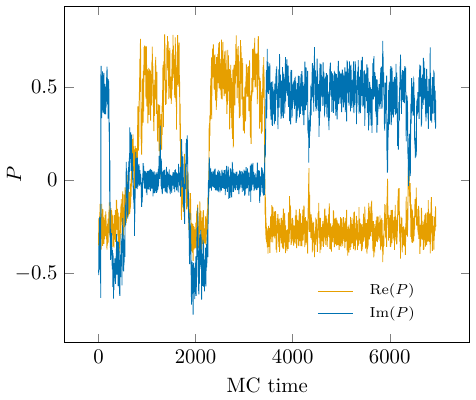}
 \caption{
 Monte Carlo history of the real (gold) and imaginary (blue) part of the Polyakov loop on a $16^3\times 6$ ensemble at high temperature. Frequent jumps between the center sectors are observed.
 \label{fig:ploop_hist1}
 }
 \bigskip
 \includegraphics[width=0.9\columnwidth]{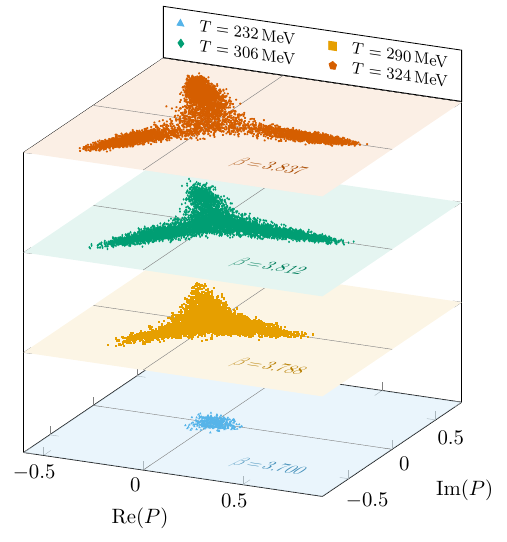}
 \caption{
 Scatter plot of the Polyakov loop in the complex plane on $16^3\times6$ ensembles at different temperatures.
 \label{fig:ploop_hist2}
 }
\end{figure}

To set the stage, we first analyze the Monte Carlo history of the unrotated Polyakov loop~\labelcref{eq:Pdef}. \Cref{fig:ploop_hist1} shows the results for our $16^3\times6$ ensemble at $\beta=3.85$, which lies deep in the high-temperature region. Along the Monte Carlo trajectory, frequent jumps occur between the three center sectors, indicating the realization of the $\ZT$ symmetry and the ergodicity of our algorithm. The symmetry of the distribution of $P$ in the complex plane is visualized in \cref{fig:ploop_hist2}. By comparing different values of $\beta$, the sharp change in the distribution is revealed as the spontaneous breaking of the $\mathrm{Z}(3)$ symmetry at high $T$ unfolds.

\begin{figure}[tb]
 \centering
 \includegraphics[height=0.275\textheight]{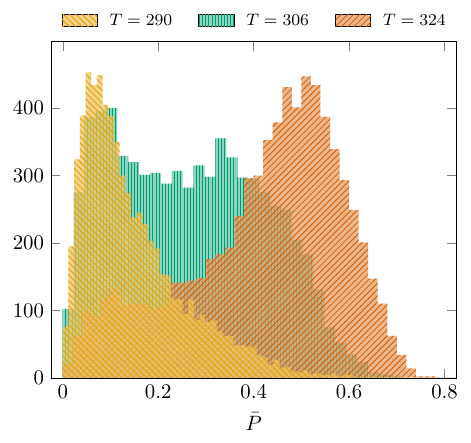}
 \caption{
 Histogram of the rotated Polyakov loop on our $16^3\times6$ ensembles, with the same color coding for the temperatures as in \cref{fig:ploop_hist2}.
 \label{fig:Pbar_hist1}
 }
 \bigskip
 \centering
 \includegraphics[width=0.9\columnwidth]{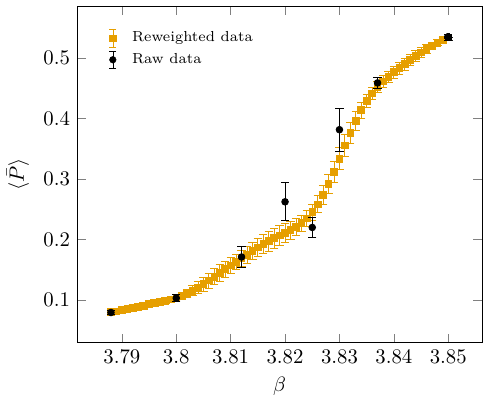}
 \caption{
 Temperature-dependence of the rotated Polyakov loop expectation value, together with an interpolation obtained using the multi-histogram method for our $20^3\times6$ ensemble.
 \label{fig:Pbar_hist2}
 }
\end{figure}

Next, we turn to the rotated Polyakov loop~\labelcref{eq:Pbardef} and present its histogram in \cref{fig:Pbar_hist1} for the same lattice ensemble. The three curves correspond to $\beta=3.788$, $\beta=3.812$, and $\beta=3.837$, using the same color coding as in \cref{fig:ploop_hist2}. In this representation, the nature of the confinement-deconfinement transition becomes more apparent: a double-peak structure is forming near the critical temperature (green), connecting the confined (yellow) and deconfined (red) distributions. To be more quantitative, we will study higher moments of this distribution. To achieve that, our data for the Polyakov loop is interpolated using the multi-histogram method~\cite{Ferrenberg:1989ui,newmanb99} discussed in \cref{sec:mh}. An example for the Polyakov loop expectation value is shown in \cref{fig:Pbar_hist2}, showcasing the effective interpolation taking into account all simulation points.

\begin{figure*}[tb]
 \centering
 \includegraphics[width=0.9\textwidth]{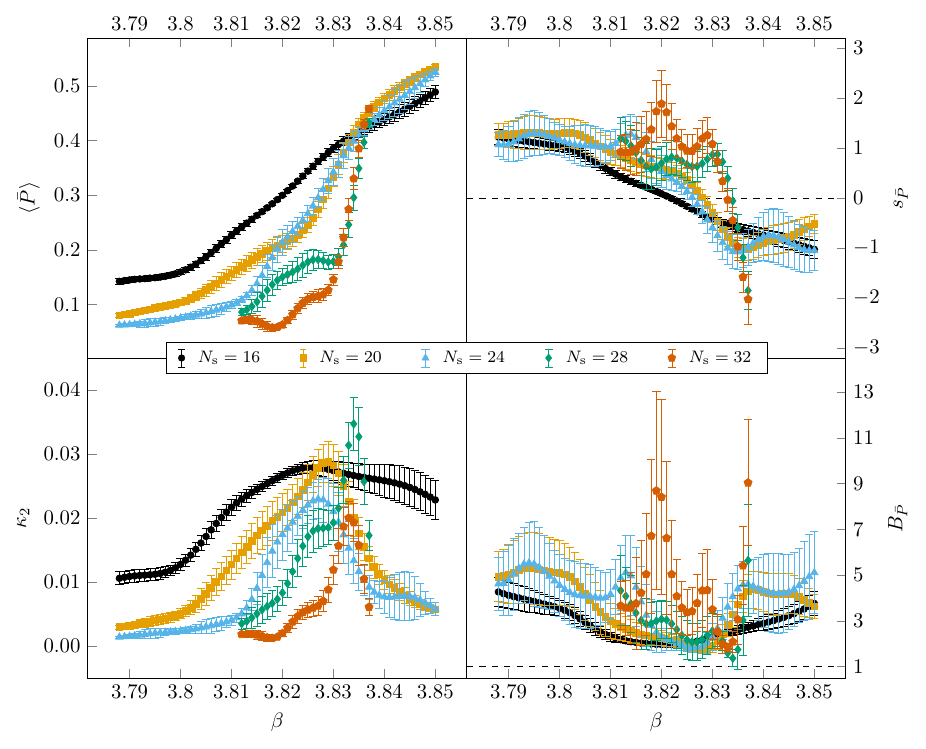}
 \caption{
 Moments of the distribution of the rotated Polyakov loop as a function of $\beta$ for different volumes. The panels show $\langle \bar P\rangle$ (top left), $\kappa_2$ (bottom left), $s_{\bar P}$ (top right) and $B_{\bar P}$ (bottom right).
 \label{fig:moments}
 }
\end{figure*}

Using the multi-histogram method, the interpolation of the higher moments~\labelcref{eq:Pbarmoments} is also straightforward. In \cref{fig:moments}, we display the $\beta$-dependence of the first four moments: the mean, the variance, the skewness and the kurtosis of the distribution of the rotated Polyakov loop. Here we include all five volumes, ranging between $16^3\times 6$ and $32^3\times 6$.

The expectation value $\langle\bar P\rangle$ shows that, in the confined regime, the observable rapidly converges toward zero, while the change across the transition becomes increasingly abrupt as the spatial volume grows, providing a strong hint that the transition is not a crossover like at zero chemical potential. The variance $\kappa_2$ likewise demonstrates behavior characteristic of a first-order phase transition: its peak becomes progressively narrower with increasing $V$. Notice that the Polyakov loop susceptibility is related to $\kappa_2$ through \cref{eq:Pbarmoments}. Thus, the approximate volume-independence of the peak height of $\kappa_2$ implies that the susceptibility satisfies $\chi_{\bar P}\propto V$ at its maximum. This is the expected behavior for a first-order transition.

The skewness of the $\bar P$ distribution also exhibits an increasingly sharp dependence, crossing zero at \mbox{$\beta_c=3.835(2)$}. This is our definition for the critical coupling.\footnote{We note that an alternative definition for the critical coupling may also be given directly in terms of the histogram of $P$~\cite{Borgs:1990ch,Borgs:1991en,Berg:2013jna,Francis:2015lha}.} Regarding the kurtosis, we observe sizeable statistical uncertainties and therefore cannot draw firm conclusions about its behavior, except that its value $B_{\bar P}(\beta_c)$ in the thermodynamic limit is compatible with unity, the value expected for a first-order phase transition. Altogether, these results give compelling evidence that the deconfinement transition in this system is indeed of first order.

\subsection{Quark condensate}

\begin{figure}[tb]
 \centering
 \includegraphics[width=0.9\columnwidth]{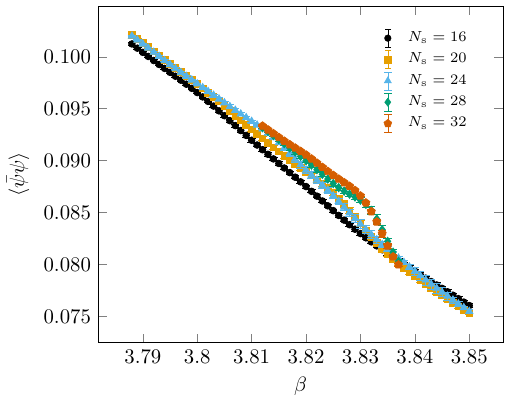}
 \caption{
 Expectation value of the quark condensate as a function of $\beta$ for different volumes.
 \label{fig:cond1}
 }
 \bigskip
  \includegraphics[width=0.9\columnwidth]{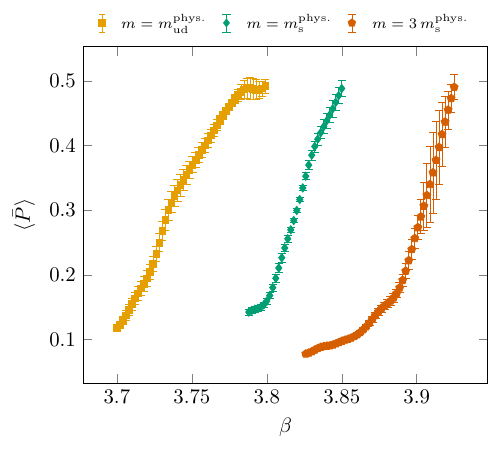}
 \caption{
 The $\beta$-dependence of the rotated Polyakov loop on the $16^3\times6$ ensemble for three different quark masses.
 \label{fig:cond2}
 }
\end{figure}

We proceed with the analysis of the quark condensate. 
\Cref{fig:cond1} displays $\langle \bar\psi\psi\rangle$, interpolated in $\beta$ via the multi-histogram procedure for several lattice volumes. Here we do not carry out the renormalization of this observable, yet we clearly observe the imprints of the abrupt change of the Polyakov loop background around the critical $\beta$, where a discontinuity of the condensate forms as $V\to\infty$.

At this stage it is natural to ask whether the observed behavior could be interpreted in terms of chiral symmetry restoration. Up to this point, our study has been performed using the physical strange quark mass, $m=m_s^{\rm phys}$. However, the emergence of a first-order transition in this setup is expected to persist for any value of $m$, provided that all three flavors share the same mass and the imaginary chemical potentials follow the prescription of \cref{eq:imagiso_choice}. To examine the quantitative mass dependence, we conducted an additional scan on the $16^3\times6$ lattice at two further mass values: the physical light-quark mass, $m=m_{ud}^{\rm phys}=m_s^{\rm phys}/28.15$~\cite{Borsanyi:2010cj}, and a heavier mass, $m=3\cdot m_s^{\rm phys}$. The resulting $\beta$-dependence of the rotated Polyakov loop is shown in \cref{fig:cond2}. While the qualitative features of the transition are similar across all mass values, a clear dependence of the critical coupling on the quark mass, $\beta_c(m)$, becomes evident. In particular, the critical $\beta$ values are read off by requiring that $\langle\bar P\rangle$ takes the same value as at the critical temperature in our finite size scaling analysis above.

Recall that to convert the $\beta$ values to lattice spacings, i.e.\ to temperature values, we have been using the lattice scale determined at the physical point. In order to compare the critical temperatures for theories that differ by such large quark mass differences, it is instructive to consider dimensionless combinations. To that end, we consider the $w_0$ scale~\cite{BMW:2012hcm} determined from the Wilson flow~\cite{Luscher:2010iy}. We calculated $w_0/a$ on $24^4$ lattices at zero chemical potential and at the critical inverse gauge couplings determined from \cref{fig:cond2}. We list the values for $T_cw_0$ in \cref{tab:Tcw0}. The results we find are remarkably close to the value $T_cw_0=0.2507(2)$ in pure gauge theory~\cite{Borsanyi:2021yoz} but deviate substantially from the approximate value $T_cw_0\approx0.14$ that this combination takes at the crossover transition in physical $(2+1)$-flavor QCD~\cite{BMW:2012hcm}.

\begin{table}[tb]
	\centering
	\setlength{\tabcolsep}{10pt} 
	\begin{tabular}{*{4}{c}}
		\toprule
		 $m_{u,d,s}$ & $m_{ud}^{\rm phys}$ & $m_s^{\rm phys}$ & $3\cdot m_s^{\rm phys}$ \\
		 \midrule
		 $T_cw_0$ & 0.239(1) & 0.248(1) & 0.2505(5)  \\
		 \bottomrule
	\end{tabular}
	\caption{Measurements of the $w_0$ scale on $24^4$ lattices at $\mu_f=0$ at the critical couplings inferred from finite temperature $16^3\times6$ ensembles. \label{tab:Tcw0}}
\end{table}

\begin{figure}[tb]
 \centering
 \mbox{
 \includegraphics[width=0.99\columnwidth]{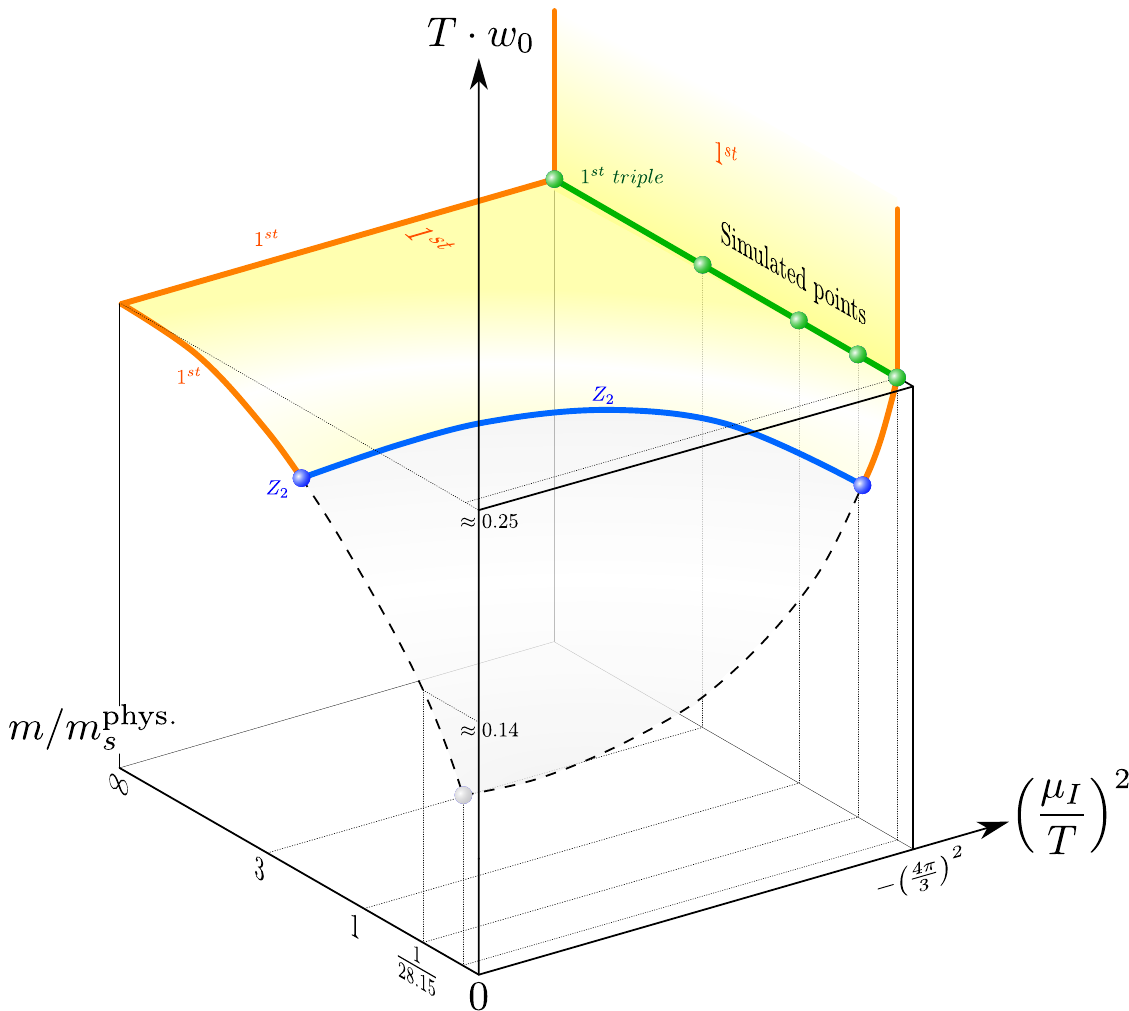}
 }
 \caption{
 Phase diagram of three-flavor QCD in the space spanned by the temperature normalized by $w_0$, the (squared) isospin chemical potential and the (degenerate) quark mass. The yellow area and orange lines indicate first-order phase transitions, the blue line a second-order transition in the $\mathrm{Z}(2)$ universality class and the dashed lines crossovers. The vertical yellow area corresponds to a first-order transition separating deconfined phases with different Polyakov loop sectors, while the mostly horizontal area denotes the transition between confined and deconfined phases. Thus, the green line constitutes a line of first-order triple points, on which our simulation points also lie (green dots).
 \label{fig:phasediag}
 }
\end{figure}

Putting together the above findings, we are able to sketch the phase diagram of the theory. This is shown in \cref{fig:phasediag} in the three-dimensional space spanned by the temperature, the isospin chemical potential and the quark mass. The system exhibits the exact center symmetry on both back-planes of this phase diagram: the one at $\mu_I/T=i4\pi/3$ as well as the one at $m=\infty$.
The first-order deconfinement phase transition that we simulated above at $\mu_I/T=i4\pi /3$ and $m=m_s^{\rm phys}$ -- and expect to persist for practically any nonzero quark mass -- is therefore continuously connected to the first-order region at $\mu_I=0$ and high quark masses. The latter is customarily depicted as the section of the diagonal in the upper right corner of the standard Columbia plot, see e.g.\ \refcite{Aarts:2023vsf}. This first-order transition terminates in a $\mathrm{Z}(2)$ second-order point, from which a $\mathrm{Z}(2)$ line emanates in \cref{fig:phasediag}. For even lower quark masses, the transition is an analytic crossover. Following \cref{tab:Tcw0} and the above quoted values, we have also indicated the qualitative behavior of the transition temperature (in units of $1/w_0$) in the figure.

\Cref{fig:phasediag} exhibits yet another region of first-order phase transitions, corresponding to the upper part of the $\mu_I/T=i4\pi/3$ back-plane, marked yellow.
That is, at high temperatures above deconfinement, there is a transition as $\mu_I$ passes from below to above $4\pi i T/3$.
For $\mu_I < 4\pi iT/3$ the real sector dominates, and for $\mu_I > 4\pi iT/3$ the two complex sectors dominate and are equivalent, and the $Z_2$ symmetry between these sectors is spontaneously broken.
Turning back to the perturbative phase diagram in \cref{fig:pertpd}, this transition corresponds to the change from the black region to the blue/yellow boundary region along the upper-left to lower-right diagonal.
We note finally that the analytical continuation of this phase diagram to real values of $\mu_I$ is known to exhibit further phases involving the Bose-Einstein condensation of charged pions, encoded in the chiral behavior of the theory. This region of the phase diagram has been studied for physical~\cite{Brandt:2017oyy,Brandt:2018omg} as well as for lighter-than-physical quark masses~\cite{Brandt:2025tkg,Brandt:2025ywy}.
\vspace*{-2mm}

\section{Conclusions}

In this paper, we discussed QCD with three mass-degenerate quarks at imaginary chemical potentials. Using the one-loop perturbative effective potential for the Polyakov loop, we mapped out the Roberge-Weiss phase transitions and the associated phase diagram of the theory in the high temperature limit. We determined the manifold, where the system exhibits an exact center symmetry and specialized to one point on it: $\mu_B=\mu_S=0$ and $\mu_I=i4\pi T/3$. Using  lattice simulations with dynamical stout-improved staggered quarks at one lattice spacing and a dedicated multi-histogram analysis, we demonstrated that the deconfinement phase transition in this setting is first order. The corresponding discontinuity in the Polyakov loop was shown to leave its imprint in the quark condensate as well. Finally, we sketched the qualitative behavior of the QCD phase diagram in the mass-chemical potential-temperature space. Our results constitute a non-trivial example, beyond pure gauge theory, for a center-symmetric quantum field theory with a real phase transition.

\begin{acknowledgments}
This research was funded by the DFG (Collaborative Research Center CRC-TR 211 ``Strong-interaction matter under
extreme conditions''~--~project number \mbox{315477589~--~TRR 211}). GE received funding from the Hungarian National Research, Development and Innovation Office~--~NKFIH (Research Grant Hungary 150241) and the European Research Council (Consolidator Grant 101125637 CoStaMM).
Views and opinions expressed are however those of the authors only and do not necessarily reflect those of the European Union or the European Research Council. Neither the European Union nor the granting authority can be held responsible for them.
\end{acknowledgments}


\appendix*
\def\thesection{\texorpdfstring{\unskip}{}}%
\makeatletter
\def\theequation@prefix{A}%
\makeatother

\section{Perturbative Phase Structure at High Temperatures}
\label{app:pert}

Expanding around a homogeneous gluon field background in the gauge where $A_4$ is time-independent,
the one-loop effective potential of QCD with $N_{\text{f}}$ massless flavors depends on the three eigenvalues $C_i$ of the temporal
component $A_4$ of the gauge bosons, as well as the angles $\theta_f=i\mu_f/T$ (the imaginary chemical potential values are effectively angles which shift the boundary conditions for each quark species $f\in\{ 1\dots N_\text{f} \}$ \cite{PhysRevD.24.475,PhysRevD.25.2667,Roberge:1986mm}).
Specifically,
\begin{align}
    V_{\text{eff}}^{\text{tot}}(  C_i , \theta_f )
    & = V_{\text{eff}}^{\text{glue}}(  C_i ) \notag \\
    & + \sum_{f=1}^{N_f} V_{\text{eff}}^{\text{1-fermion}}( C_i + \theta_f T )\,.
\end{align}
Notice that $V_{\text{eff}}^{\text{tot}}$ depends on \textsl{all} of the $C_i$ and $\theta_f$, while each 1-fermion potential depends on all $C_i$, which are uniformly shifted by a specific $\theta_f T$.

In the one-loop approximation, we can give a closed-form expression for the two pieces of the effective potential:
\begin{align}
    &V_{\text{eff}}^{\text{glue}}\left( C_i  \right) \\ \notag
    &= \frac{\pi^2 T^4}{24} \sum_{j,k=1}^N\left(1-\left( \left[\frac{C_j}{\pi T}-\frac{ C_k}{\pi T }\right]_{\!\bmod 2}-1\right)^2\right)^2
\\
    &V_{\text{eff}}^{\text{1-fermion}} \left(C_i \right)\\ \notag
    &=-\frac{\pi^2 T^4}{12} \sum_{i=1}^N \left( 1- \left( \left[ \frac{ C_i}{\pi T} +1 \right]_{\!\bmod 2} -1 \right)^2 \right)^2,
\end{align}
omitting terms that do not depend on the background field value $C_i$.
Here $[{}\dots{}]_{\!\bmod2}$ indicates that the quantity is to be shifted by an even integer to fall into the range $\left[0,2\right)$.

Obtaining the expectation value of the background field $C_i$ is equivalent to determining the minimum of the effective potential. At high temperatures, the one-loop effective potential is approximately equal to the full effective potential. The fact that we are analyzing the high temperature regime also justifies considering massless quarks. 

The minima candidates sit at discrete points $\left( C_1,\dots,C_{N-1},C_{N}\right)=\left(\phi,\dots,\phi, -(N-1)~ \phi \right)$, where
$\phi=a_n=2\pi T n / N$
for a general number $N$ of colors. 
Those are the exact minima of the pure Yang-Mills effective potential $V_{\text{eff}}^{\text{glue}}(C_i)$, indexed by $n \in \left \{0, \dots, N-1  \right \}$. Note that the last entry is fixed via the tracelessness condition of the Lie-algebra. Setting these points as the minimum candidates of the total effective potential is an assumption which should hold true as long as the number of fermion flavors is not too large compared to the rank of the gauge group%
\footnote{%
For example, setting $N_\text{f}=10$ and $N=2$, we find that if 5 quarks have $\theta=0$ and five quarks have $\theta=\pi$, then the choice $(C_1,C_2)=(\pi T/2,3\pi T/2)$ is preferred over either conventional center-symmetric minimum.
A similar counter-example for $N=3$ is $N_\text{f}=24$ with 12 flavors at $\theta=13\pi/27$ and 12 flavors at $\theta=\pi/3$, albeit out of the range of asymptotic freedom.
}.
Not only does this reduce the minimization procedure from an $(N-1)$-dimensional problem to a one-dimensional one, but one can see that the remaining parameter $\phi$ ends up being the angle of the Polyakov loop.

Specifying to the case $N_{\text{f}}=3$ and $N=3$ at a fixed temperature $T$, we will find a three-dimensional phase diagram, where each phase corresponds to a different value $a_n$ of the Polyakov loop angle $\phi$. We can consider this phase diagram in the $\theta_u$, $\theta_d$, $\theta_s$ basis, or, equivalently, in the $\theta_B$, $\theta_I$, $\theta_S$ basis, as defined in \cref{eq:chempotbasis} for the chemical potentials. 
As we pointed out already in footnote~\ref{fn:1}, the latter parameterization is a double cover of the former one. To see this, we write down the transformation between the two coordinate systems via
\begin{equation}
\begin{pmatrix}
\theta_u \\
\theta_d \\
\theta_s
\end{pmatrix}
=
    U\!
\begin{pmatrix}
\theta_{I} \\
\theta_{B} \\
\theta_{S}
\end{pmatrix}
\ \ \text{with} \ \  U=\begin{pmatrix}
1/2 & 1/3 & 0 \\
-1/2 & 1/3 & 0\\
0 & 1/3 & -1
\end{pmatrix}.
\end{equation}
Computing the determinant of the transformation matrix, we find $\det [U]=1/3$. However, comparing the volumes of the two parameter spaces, we find $6 \pi \cdot 4 \pi \cdot 2 \pi= 6 \cdot (2 \pi)^3$, i.e. the parameter space of the $\left(\theta_{I}, \theta_{B}, \theta_{S} \right)$-basis is six times as large as the that of the $\left(\theta_{u}, \theta_{d}, \theta_{s} \right)$-basis, hence we cover twice the volume in chemical potential space. Thus, for instance $(\theta_B,\theta_I,\theta_S)=(3\pi,2\pi,\pi)$ is equivalent to $(\theta_B,\theta_I,\theta_S)=(0,0,0)$.

\begin{figure}[t]
  \centering
  \includegraphics[width=0.32\columnwidth]{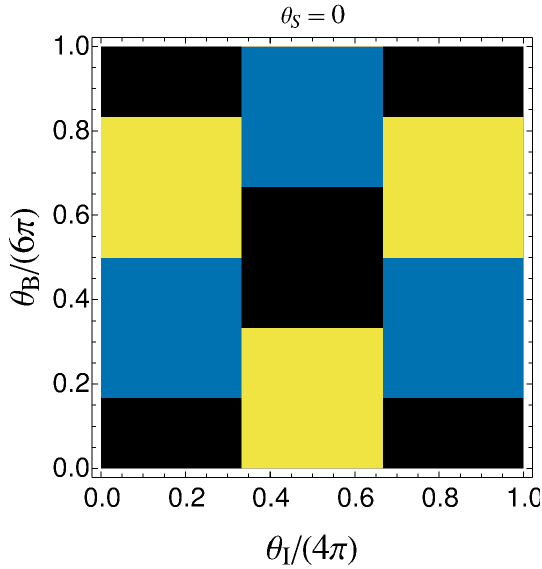}\hfill
  \includegraphics[width=0.32\columnwidth]{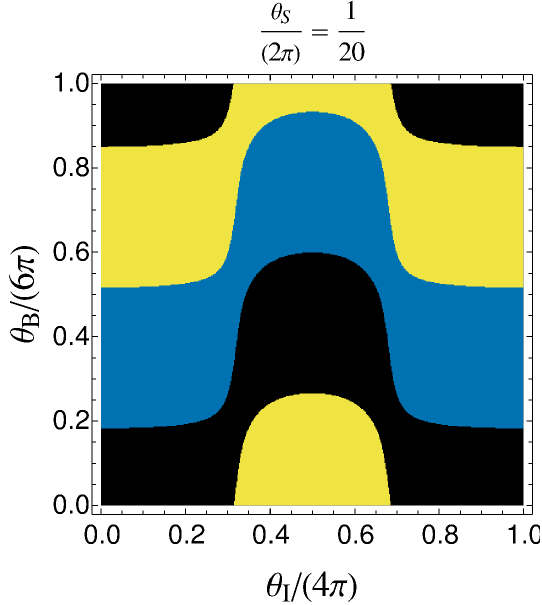}\hfill
  \includegraphics[width=0.32\columnwidth]{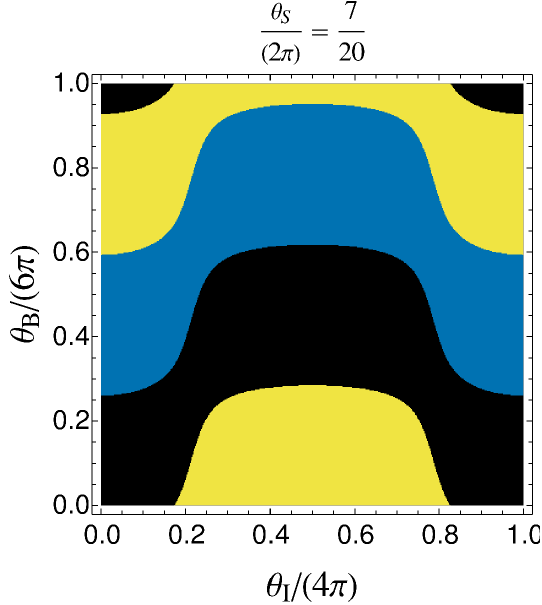}\hfill
  \includegraphics[width=0.32\columnwidth]{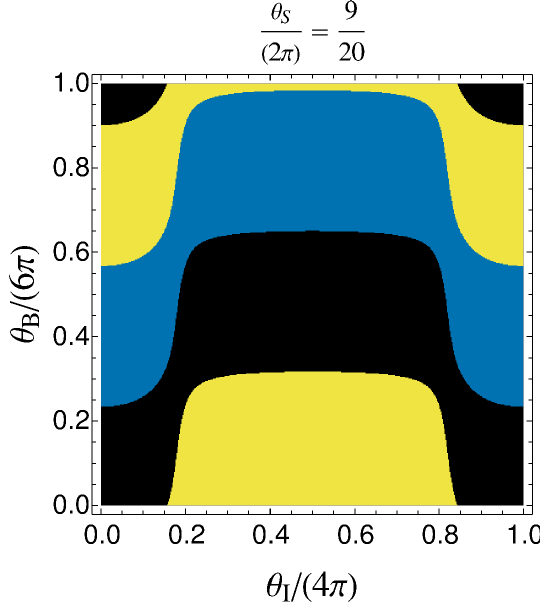}\hfill
  \includegraphics[width=0.32\columnwidth]{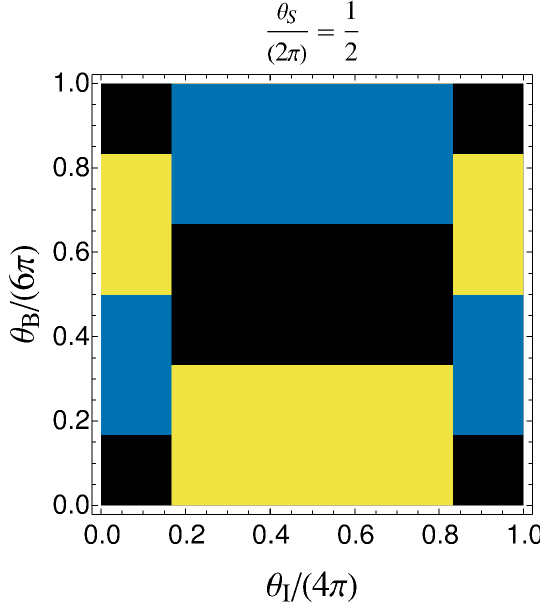}\hfill
  \includegraphics[width=0.32\columnwidth]{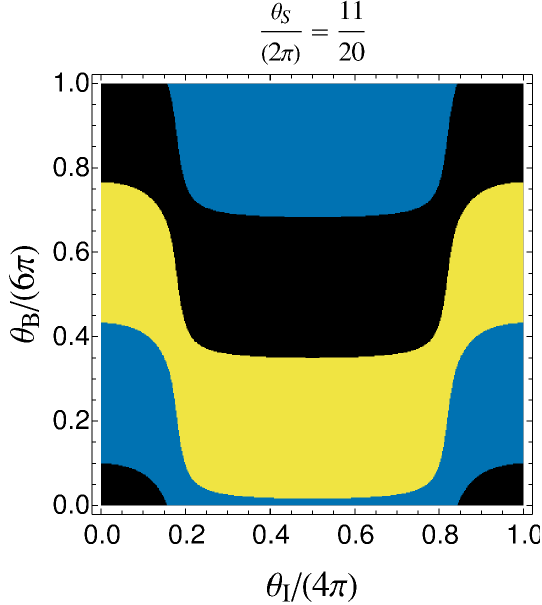}\hfill 
  \includegraphics[width=0.32\columnwidth]{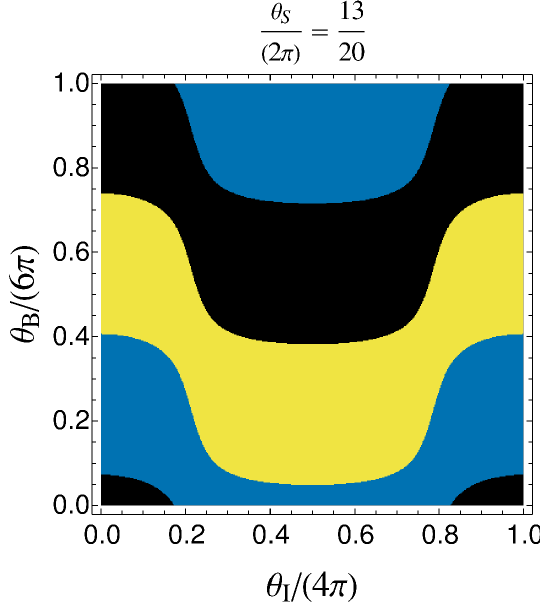}\hfill
  \includegraphics[width=0.32\columnwidth]{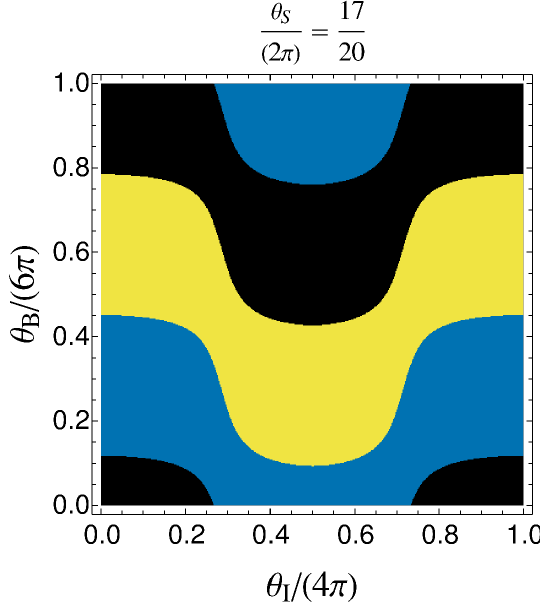}\hfill 
  \includegraphics[width=0.32\columnwidth]{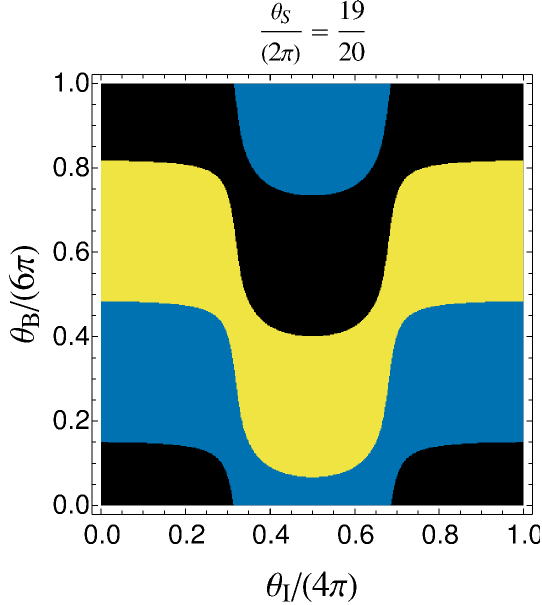}
  \caption{Phase diagrams for three flavors evaluated at a constant value of the strangeness chemical potential $\theta_{S}$ in the parameterization \cref{eq:chempotbasis}. At $\theta_{S}=0$, we can observe that the $\theta_{I}=0$-line reproduces the one-flavor case.}
  \label{fig:chemPotBasisSlices}
\end{figure}

We first consider the $\theta_B$, $\theta_I$, $\theta_S$ basis and numerically compute two-dimensional slices of the full phase diagram where $\theta_S$ is constant, see \cref{fig:chemPotBasisSlices}. 
As one can observe in the definition, the three-flavor case differs from the two-flavor case by one additional fermionic contribution to the effective potential at the fixed value $\theta_s$, which needs to be accounted for in the minimization procedure. More explicitly:
\begin{align}
    &V_{\text{eff}}^{\text{tot}}(C_i,\theta_f)
    \\ \notag
    &=V_{\text{eff,2-flavor}}^{\text{tot}}(C_i,\theta_u,\theta_d)
    + V_{\text{eff}}^{\text{1-fermion}}(C_i + \theta_{s} T).
\end{align}
Interestingly, this additional term drastically changes the qualitative structure of the overall phase diagrams. While the two-flavor diagram has straight phase boundaries forming a ``honeycomb'' shape, as observed both in perturbative and non-perturbative studies~\cite{Brandt:2022jwo}, the three-flavor diagrams have curved phase boundaries, with measure zero exceptions.

\begin{figure}[tb]
\centering
\includegraphics[width=0.47\columnwidth]{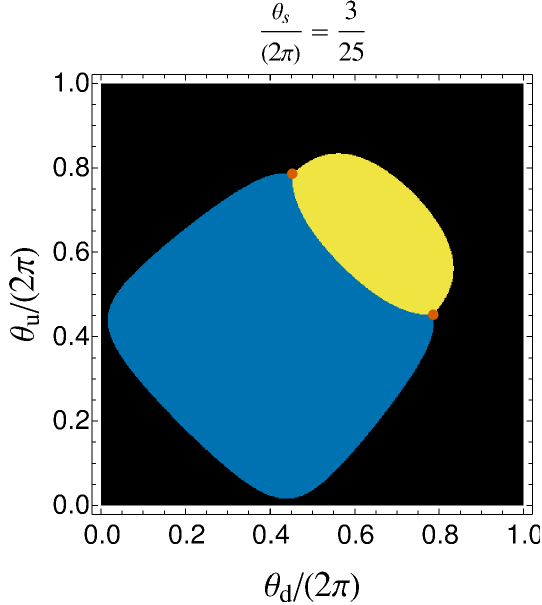}\hfill
\includegraphics[width=0.47\columnwidth]{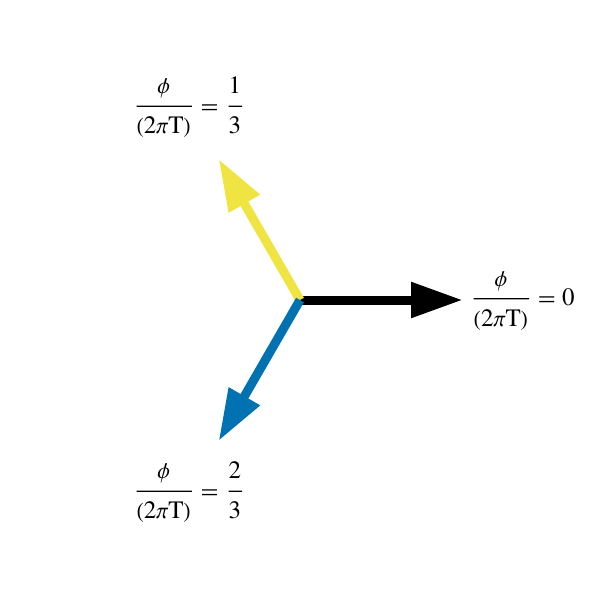}\hfill
\includegraphics[width=0.47\columnwidth]{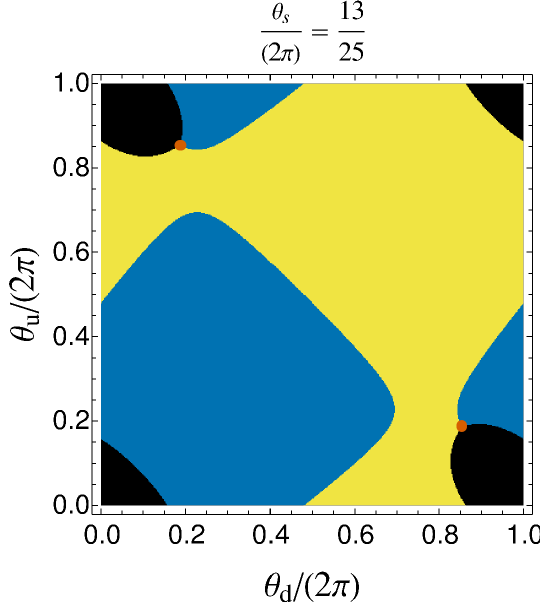}\hfill
\includegraphics[width=0.47\columnwidth]{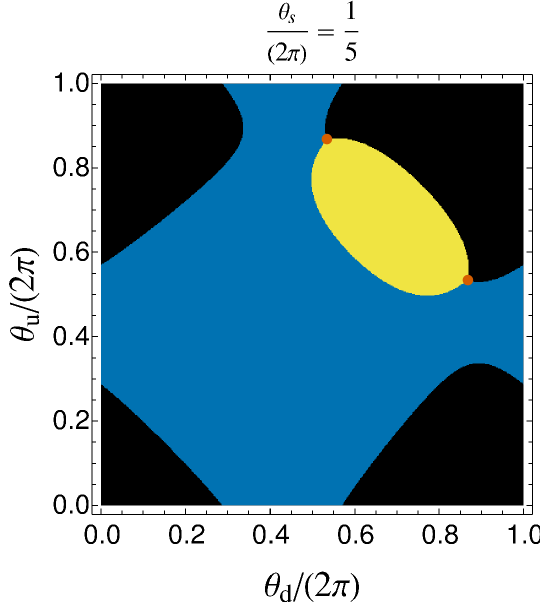}
\caption{Phase diagrams for three flavors evaluated at various (constant) values of the strange quark chemical potential $\theta_{s}$, representing the typical structure of these diagrams. The red dots indicate triple points. The arrows in the top right panel visualize the meaning of the three colors as Polyakov loop angles.}
\label{fig:ExampleSlicesPerturbationTheoryV1}
\end{figure}

An interesting observation is that the two vertical lines in the $\theta_{S}=0$-slice and $\theta_{S}/\left(2 \pi \right)=1/2$-slice in \cref{fig:chemPotBasisSlices} are not simple phase transition lines between two phases but, in fact, lines of triple points. This fact gets somewhat obscured in this coordinate system, since the third phase ``hides'' in a slice with a slightly different value of $\theta_{S}$.
Changing the basis to the up-, down-, strange-quark chemical potentials $\left( \theta_u,\theta_d,\theta_s \right) \in \left[0,2 \pi \right)^3$ makes the triple points immediately visible. Keeping $\theta_s$ fixed, a collection of slices is displayed in \cref{fig:ExampleSlicesPerturbationTheoryV1}. These three slices exemplify the typical structure of these phase diagrams. 

\begin{figure}[t]
  \centering
  \includegraphics[width=0.32\textwidth]{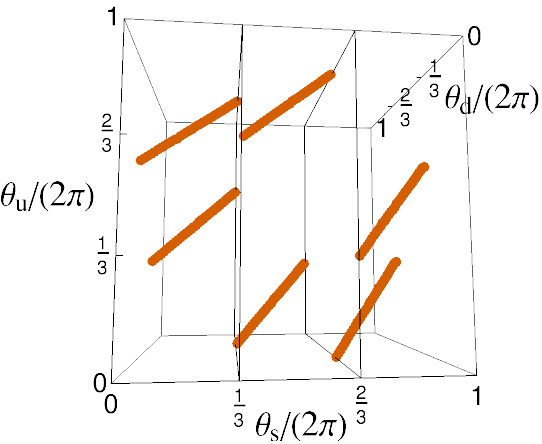}
  \caption{Triple points in the high-temperature phase diagram of three-flavor QCD, generated numerically.
  }
  \label{fig:TriplePointsPerturbationTheory}
  \bigskip
  \includegraphics[width=0.32\columnwidth]{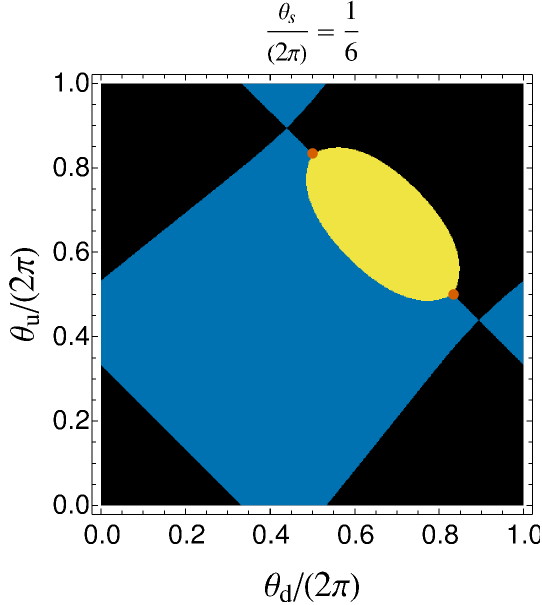}\hfill
  \includegraphics[width=0.32\columnwidth]{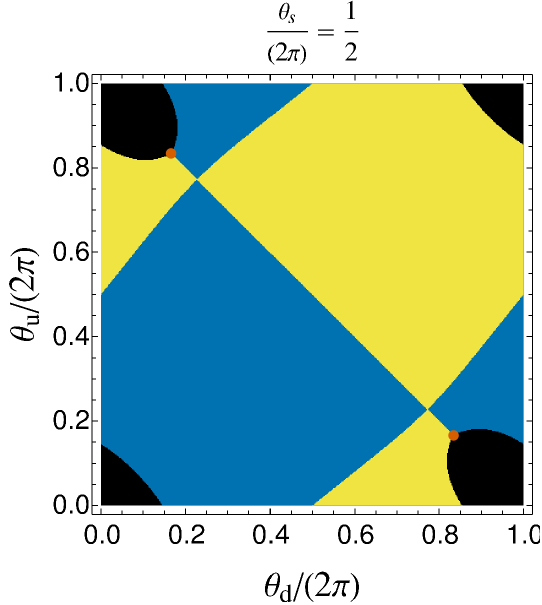}\hfill
  \includegraphics[width=0.32\columnwidth]{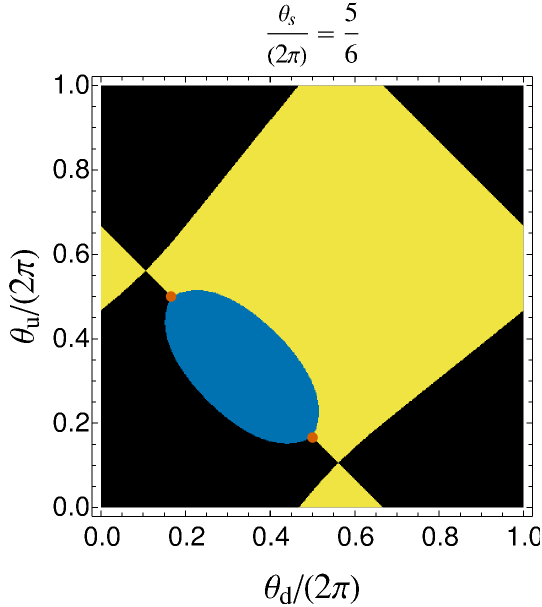}\hfill
  \includegraphics[width=0.32\columnwidth]{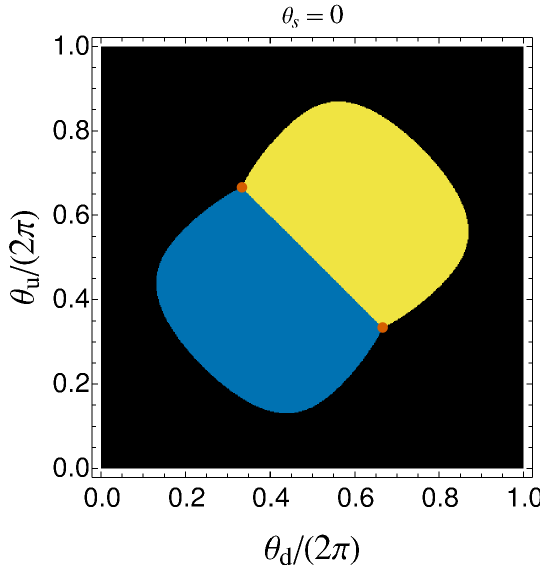}\hfill
  \includegraphics[width=0.32\columnwidth]{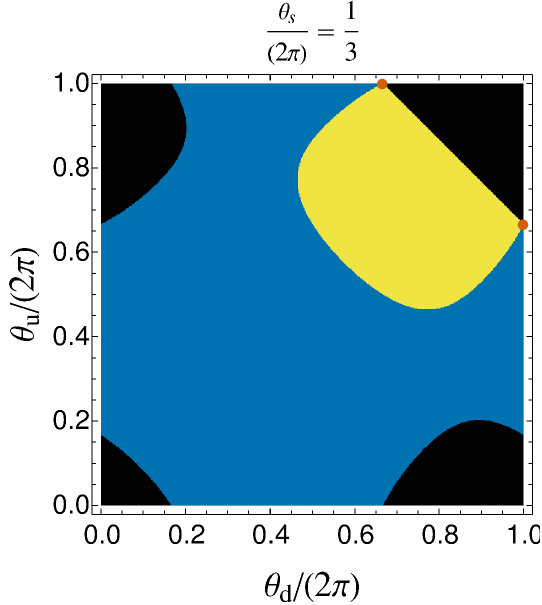}\hfill
  \includegraphics[width=0.32\columnwidth]{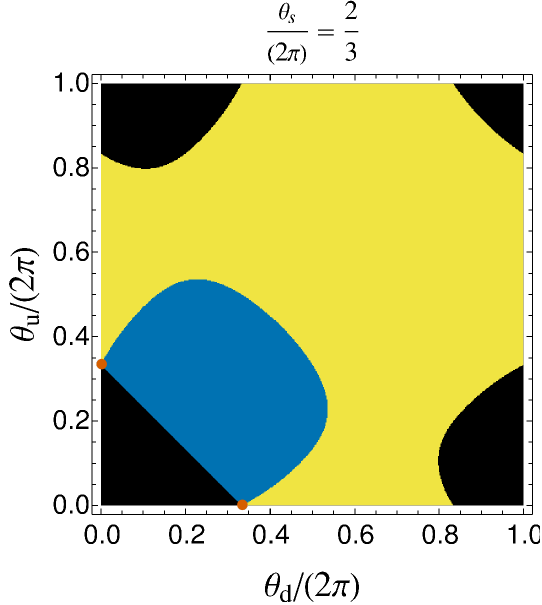}
  \caption{Phase diagrams for three flavors evaluated at constant strange quark chemical potential $\theta_{s}$ taking values at phase transition points in the one-flavor theory (upper row) and points in between two transition points  (lower row). The red dots indicate triple points.}
  \label{fig:SpecialSlicesPerturbationTheory}
\end{figure}

We can see that in this basis, each fixed-$\theta_s$ slice has exactly two triple points. These triple points move linearly with $\theta_s$, drawing a trajectory  which can be plotted
in the full, three-dimensional phase diagram, see \cref{fig:TriplePointsPerturbationTheory}. 
The trajectories of the triple points can be parameterized as
\begin{equation}  
 \theta_s = \left[\theta_u \mp \frac{2\pi}{3}\right]_{\bmod{2 \pi}} = \left[\theta_d \pm \frac{2\pi}{3}\right]_{\bmod{2 \pi}}.
\end{equation}
Plugging this trajectory into the effective potential we can verify that 
\begin{align}
    V_{\text{eff}}^{\text{tot}}(\phi=a_0,\theta_f^{T}) & = 
    V_{\text{eff}}^{\text{tot}}(\phi=a_1,\theta_f^{T}) \notag\\
    & = V_{\text{eff}}^{\text{tot}}(\phi=a_2,\theta_f^{T})
\end{align}
which is the definition of a triple point. The triple points realize the exact $\ZT$ symmetry that we pointed out before \cref{eq:imagiso_choice} of the main text. Performing a coordinate transform, one can verify that these two linear trajectories coincide with two of the four linear triple-point lines in the $\theta_B$, $\theta_I$, $\theta_S$ coordinate system, visible in \cref{fig:chemPotBasisSlices}. The other two triple-point lines are a consequence of the double-cover feature discussed earlier.

Keeping in mind that the third fermion flavor acts as an additional fermionic term in the effective potential,  it is intuitive that something interesting might happen when we set $\theta_s$ to one of the values where the one-flavor theory would have its phase transition, i.e. $\theta_{s}/(2 \pi) \in \left \{ 1 /6, \ 1/2 , \ 5 /6  \right \}$.
As one can observe in \cref{fig:SpecialSlicesPerturbationTheory}, we now find both straight and curved phase transition lines. This can be traced back to the symmetry properties of the mod-function. The points right between two of these phase transition values also satisfy this symmetry property, causing the phase diagrams at these points to also have an atypical structure. 
The $\theta_s=0$ slice of this figure was shown already in \cref{fig:pertpd} of the main text.
Further details of the effective potential have been studied in \refcite{Adam_thesis}.

\FloatBarrier
\clearpage
\newpage

\bibliographystyle{JHEP}
\bibliography{3fl.bib}

\end{document}